\documentclass[prl,aps,twocolumn,showpacs,floatfix,floats,nobalancelastpage]
{revtex4}
\usepackage{graphicx,bm} 	%\usepackage{amsmath,amssymb,epsf,bm}

\begin{document}
%\draft
\title{Spin transmission through quantum dots with strong 
spin-orbit interaction}
\author{     %C.-H. Chang, A. G. Mal'shukov and K. A. Chao}
Cheng-Hung Chang$^1$, Anatoly G. Mal'shukov$^2$, and Koung-An Chao$^3$}
\date{\today}
\affiliation{
$^1$
National Center for Theoretical Sciences, Physics Division,
101, Section 2 Kuang Fu Road, Hsinchu 300, Taiwan \\
$^2$
Institute of Spectroscopy, Russian Academy of Sciences,
142190 Troitsk, Moscow Region, Russia \\
$^3$ 
Solid State Theory, Department of Physics, Lund University,
S-223 62 Lund, Sweden}

%\address{e-mail: chchang@phys.cts.nthu.edu.tw}

%%%%%%%%%%%%%%%%%%%%%%%%%%%%%%%%%%%%%%%%%%%%%%%%%%%%%%%%%%%%%%%

\begin{abstract}
Quantum oscillations of the spin conductance through  regular
and chaotic 2D quantum dots under the varying Rashba spin 
orbit interaction and at zero magnetic field have been 
numerically calculated by summing up the spin evolution matrices
for classical transmitting trajectories. Fourier analysis of these
oscillations showed power spectra strongly dependent on the 
dot geometry. For narrow rings the spectra are dominated by
a single peak in accordance with previous analytic results.
In other geometries the spectra are represented by multiple
peaks for regular QD and quasicontinuum for chaotic QD.
\end{abstract}

% 72.25.-b	Spin polarized transport
% 72.25.Dc	Spin polarized transport in semiconductors
% 73.63.-b	Electronic transport in mesoscopic or
%		nanoscale materials and structures
% 73.63.Hs	Quantum wells
% 73.63.Kv	Quantum dots
% 03.65.Vf	Phases: geometric; dynamic or topological
% 03.65.Sq	Semiclassical theories and applications

\maketitle

%\section{Introduction}

Recently spin polarized transport has attracted a considerable interest 
because of its potential in semiconductor device
applications and quantum computing \cite{Spin}.
In semiconductor quantum wells, such transport can be strongly
influenced by the spin-orbit interaction (SOI).
There are two contributions to the SOI,
the Dresselhaus term \cite{Dresselhaus} and the Rashba term \cite{Bychkov}.
The latter can be quite strong in some heterostructures \cite{Nitta}. 
There have been a number of experimental and
theoretical studies of the spin-orbit effects on the electron transport
in semiconductors.
Here, we will focus on the quasiclassic electron transport through 2D
regular and chaotic quantum dots (QD).
Special attention will be payed to the doubly connected QD geometries.
While considering such a problem, an important parameter of the characteristic
SOI length has to be introduced and compared with the QD size $L$.
This length can be defined as $L_{so}=2\hbar v_F /\Delta$, where $\Delta$
is the spin splitting of the electron energy due to the SOI and
$v_F$ is the Fermi velocity.
In Grundler's samples \cite {Nitta},
it has been found as short as 10$^{-5}$cm.
The regime of weak SOI can be realized when $L\ll L_{so}$.
In this regime, the random matrix approach has been applied \cite{Aleiner} 
to study the SOI effects on the electric transport
through a chaotic QD \cite{Folk}.
In the opposite case of the strong SOI, $L \geq L_{so}$, 
the spin dynamics in the dot becomes fast and the
random matrix approach is not helpful.
This regime is of much importance from the fundamental 
point of view and also for semiconductor device applications,
because at zero magnetic field it allows to observe periodic oscillations of transport 
parameters due to the interference of spinor amplitudes.
Their phases are accumulated under the SOI effect
during the particle motion along a quasiclassic trajectory and depend 
only on the trajectory geometry.
In QD having the ring shape constructive and destructive interference 
of the spin phases gives rise to periodic variations of the
electric conductance as a function of  the SOI strength \cite{Mathur,Aronov}. 
In practice, it is possible to change the strength of the Rashba SOI
up to $50\%$ of its value by tuning the gate voltage \cite{Nitta}.
Hence, one can observe periodic oscillations in the electron transport
by a simple gate manipulation.
%%%%%%%%%%%%%%%%%%%%%%%%%%%
\begin{figure}[htbp!]
\begin{minipage}[b]{4.2cm}
\includegraphics[width=4.1cm]{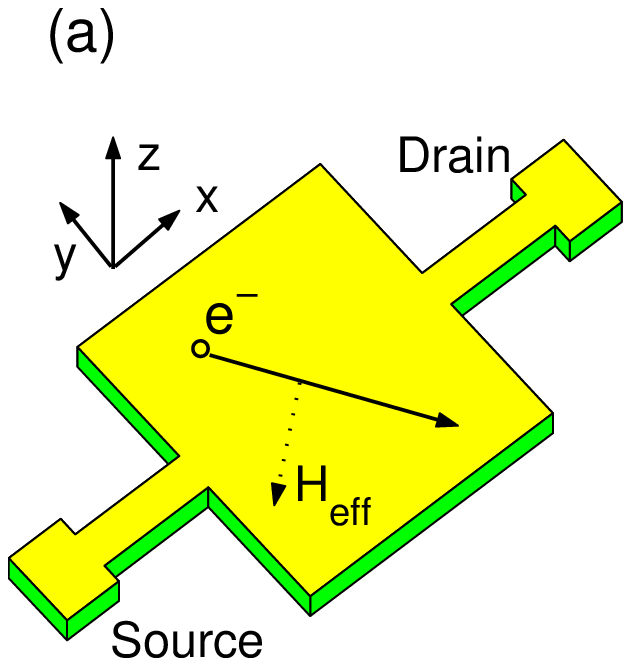}
\end{minipage}
\begin{minipage}[b]{3.3cm}
\includegraphics[width=3.2cm]{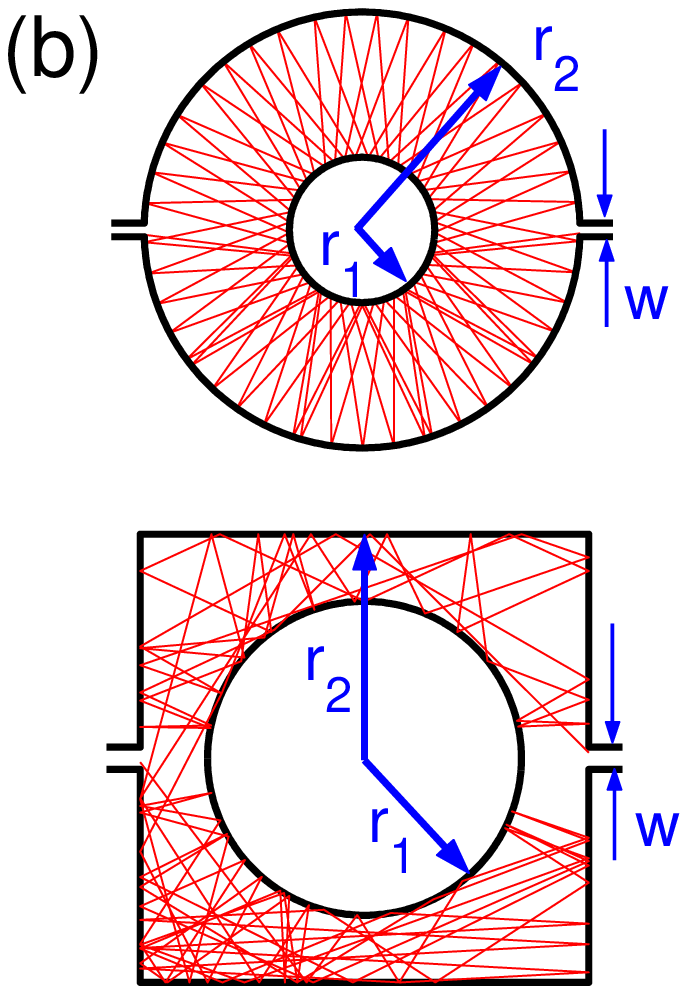}
\end{minipage}
\caption{
(a) 2D electron gas confined to a narrow layer in $z$-direction
and bounded inside a two dimensional billiard sample in $xy$-direction
is connected to the source and drain by two leads. The Rashba effect
induces the effective magnetic field $H_{\mbox{eff}}$
which is parallel to the $xy$ plane and perpendicular to the electron 
velocity direction. While an electron moves along the trajectory,
its spin precesses around $H_{\mbox{eff}}$.
(b) The geometry of the samples can be an annular billiard with the inner
and outer radius $r_1$ and $r_2$ and the lead width $w$
or the Sinai billiard with radius $r_1$ of the inner circle
and the length $2r_2$ of the outer square.}
\label{fig1}
\end{figure}
%%%%%%%%%%%%%%%%%%%%%%%%%%%

For a 2D disordered ring, the spin phase effect on the weak localization 
electric conductance \cite{Mathur}
and its pair conductance correlation function \cite{Malshukov2}
have been calculated under the assumption of a diffusive
electron propagation inside the ring.
Besides the electric transport,
the spin phase interference also affects the spin transport.
Oscillations of the mean spin conductance through a classically chaotic QD
with the shape of a thin 2D loop have been analytically calculated
\cite{Malshukov1}.
The width of the loop arms was assumed to be much less than  $L_{so}$.
In this case the particle oscillations between the loop boundaries
produce only a small effect on the spin phase.
The phase accumulates mostly when the particle advances  
along the loop arms making several windings during its escape time.
In the present work we will calculate the mean spin conductance beyond
the narrow loop geometry. 
We will employ the path integral approach and perform a numerical
simulation of the spin evolution along the classic transmission trajectories.
Weak localization corrections will be ignored under the assumption of 
a large number of transmitting channels.
The spin conductance will be calculated for regular and chaotic systems like 
the annular billiard, the square billiard, and Sinai billiard. 

An appropriate approach to the spin transport 
through a QD sample like that in Fig. \ref{fig1}
is the generalized Landauer-B\"uttiker formula
\begin{equation}\label{eq1}
g_{abcd}=\frac{1}{h}\sum_{n\,m}t^{ab}_{nm}\,t^{cd *}_{nm} ,
\end{equation}
where $t^{ab}_{nm}$ denotes the transmission amplitude of an
electron at the Fermi energy propagating  from the incoming channel $n$
to the outgoing channel $m$ and indices a and b stand for the spin variables.
The spin conductance given by Eq. (\ref{eq1}) relates the spin current
$j^{bd}_L$, flowing towards the right lead, to the spin polarization matrix
$\bm{\sigma}^{ac}\cdot\bm{N}\Delta\mu/2$ in the left lead,
where $\Delta\mu$ is the difference of the chemical potentials
between two spin projections in the left lead.
This difference gives rise to the spin polarization along the unit vector 
$\bm{N}$.
Components of $\bm{\sigma}$  are the Pauli matrices
$(\sigma_x, \sigma_y, \sigma_z)$.
The transmission amplitude can be represented as \cite{Jalabert}
$$
t^{ab}_{nm}=
-i\hbar\sqrt{v_nv_m}\int dy'\int dy \phi_n^*(y')\phi_m(y)G_{ab}(y',y),
$$
with the transverse wave function $\phi_m$ ($\phi_n$) of the incoming
(outcoming) mode and the retarded Green function $G_{ab}(y',y)$,
where $y$ ($y'$) denotes a point on the entrance (exist) cross section.
Under the semiclassical approximation, the Green function is represented by
a sum over classical trajectories propagating from one lead to the other
\cite{Jalabert}. 
Then, according to \cite{Malshukov1}, the contribution of
such a trajectory $\gamma$ to the transmission coefficient can be written as  
\begin{equation}\label{eq2}
t^{ab}_\gamma=t_\gamma S^{ab}_\gamma,
\end{equation}
where $t_\gamma$ denotes the spin independent amplitude and $S^{ab}_\gamma$
represents the $(a,b)$ matrix element
of the unitary spin evolution matrix $S_\gamma$.
In the absence of the external magnetic field, this matrix is determined
by the Rashba spin-orbit interaction and can be written as \cite{Malshukov1}
\begin{equation}\label{eq3}
S_\gamma=S_{\gamma_{n(\gamma)}}\cdots S_{\gamma_2}S_{\gamma_1},
\end{equation}
with the evolution operator $S_{\gamma_i}$
for the $i$-th straight segment of the ballistic trajectory,
where $S_{\gamma_1}$ starts from the entrance lead.
For a segment of length $l_i$ with the angle $\omega_i$ with respect to
the $x$-axis, this operator can be represented by the matrix \cite{Malshukov1}
\begin{equation}\label{eq4}
S_{\gamma_i}={\bm 1} 
\cos\left(\frac{l_i}{L_{so}}\right)
-i\sin\left(\frac{l_i}{L_{so}}\right)
\left(\sigma_y \cos\omega_i - \sigma_x \sin\omega_i\right),
\end{equation}
where ${\bm 1}$ is the  $2\times 2$ identity matrix.
The operator (\ref{eq4}) acts on spinors and
describes the spin precession around the 
axis determined by the unit vector with coordinates 
$(-\sin\omega_i, \cos\omega_i)$ through the angle $2l_i /L_{so}$.
The spin rotation length $L_{so}=\hbar/\alpha m^*$ is inversely proportional
to the Rashba coupling constant $\alpha$ and the effective mass $m^*$.
Instead of the 4-th rank spinor, which is the spin conductance in
the spinor representation in Eq. (\ref{eq1}),
it is convenient to write it as a 3D matrix 
$$g_{ij}= \sum_{abcd} \sigma^{ca}_i \sigma^{bd}_j g_{abcd},$$
which determines the spin  transport between spin orientations
$i$ and $j$.
In the configurational averaging, the spin conductance is averaged
over a small range of energy around $E_F$
or over small variations of the dot shape \cite{Jalabert}, 
which smooths out the rapid transmission oscillations  
caused by  interference of de Broigle waves.
The so averaged $\langle g_{ij}\rangle$ is then given by \cite{Malshukov1}
\begin{equation}\label{eq5}
\langle g_{ij}\rangle=\frac{1}{h}\sum_\gamma|t(\gamma)|^2\,\mbox{trace}
\{\sigma_i S_\gamma \sigma_j S_\gamma^\dagger\}.
\end{equation}

One should note a fundamental distinction of the spin phase effect,
as it shows up in the spin transport, from its effect on the charge transport. 
Indeed, the mean {\it electric} conductance is given by (\ref{eq5}) with the
trace equal to $2\delta_{ij}$.
Hence, it does not depend on the spin-orbit effects.
On the contrary, the components of the mean {\it spin} conductance oscillate,
as we will see below, due to the spin phases contained in $S_\gamma$.
It is also important to note a difference between the AB and spin phase
effects.
Unlike the latter,  
the AB effect vanishes in both the electric and spin mean conductances, 
as far as the weak localization correction is not taken into account.
This difference is a consequence of  non-Abelian nature of the spin phase. 

In the semiclassical approximation, the spin independent transmission
probability $|t(\gamma)|^2$ in (\ref{eq5}) is given 
by the transmission ratio of an ensemble of classical trajectories,
in which $|t(\gamma)|^2$ is replaced by $f_\gamma\cos(\theta_\gamma)$,
where $f_\gamma=1$ or $f_\gamma=0$ for transmitted respectively reflected
trajectories.
The value $\cos(\theta_\gamma)$ is the weight of the trajectory $\gamma$,
according to its injection angle $\theta_\gamma$ with $x$ axis \cite{Jalabert}.
The initial conditions for the trajectories are chosen such that
$n_w$ positions are uniformly distributed on the entrance
cross section with $n_\theta$ uniformly distributed injection angles 
between $-\pi/2$ and $\pi/2$ on each position.
The spin conductance is calculated
as a function of SOI strength by setting $h=1$ and taking $n_{L_{so}}$
points of $L_{so}$ between $L_{so1}$ and $L_{so2}$,
uniformly distributed on the $1/L_{so}$ axis.
%%%%%%%%%%%%%%%%%%%%%%%%%%%
\begin{figure}[htbp!]
\includegraphics[width=8.6cm]{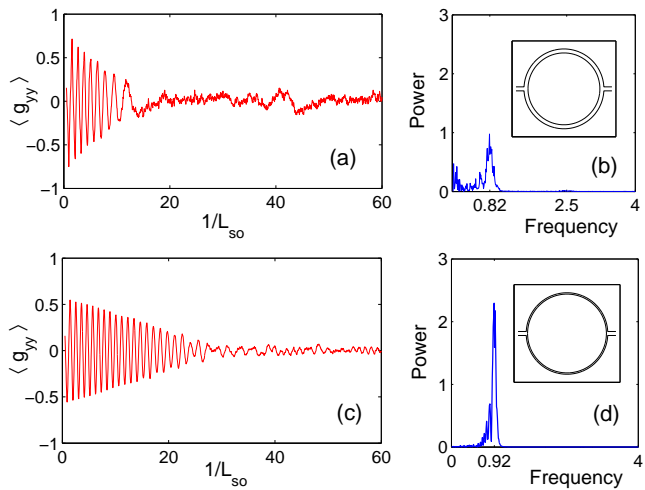}
\caption{The conductance $\langle g_{yy}\rangle$ as a function of
$1/L_{so}$ and its power spectrum in the ring billiard with
$(r_2,w,n_w,n_\theta,n_{L_{so}},L_{so1},L_{so2})$=$(1,0.1,9,80,3000,0.01,2)$,
where $r_1=0.9$ for (a), (b) and $r_1=29/30$ for (c), (d).}
\label{fig2}
\end{figure}
%%%%%%%%%%%%%%%%%%%%%%%%%%%

Figure \ref{fig2} (a) shows the conductance $\langle g_{yy}\rangle$
through the narrow ring in the inset of Fig. \ref{fig2} (b).
The conductance oscillation is regular for small $1/L_{so}$
around $1/L_{so}<10\simeq (r_2 -r_1)^{-1}$.
In this regime, the spin rotation length $L_{so}$ is larger than the
ring width and the electron spin phase does not change much between two
collisions with the sample boundary.
For $1/L_{so}>10$, the regular oscillation disappears,
because the trajectories with the same winding number arrive to
the exit lead with different  phases, 
which generate a disordered interference pattern.
The power spectrum in Fig. \ref{fig2} (b)
on the interval $0<1/L_{so}<100$ shows a dominating frequency centered
around $0.82$ with a low frequency tail.
For an extremely narrow ring like that in Fig. \ref{fig2} (d),
the regular oscillation regime is extended to around $1/L_{so}=20$.
The dominating frequency in the power spectrum is shifted to $0.92$.

\begin{figure}[htbp!]
\includegraphics[width=8.6cm]{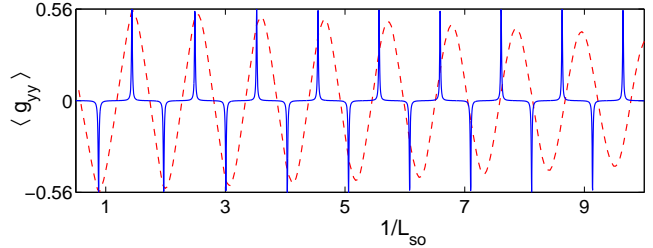}
\caption{The conductance $\langle g_{yy}\rangle$ calculated from (\ref{eq6})
with $\kappa=0.1$ (solid) and rescaled from Fig. \ref{fig2} (c) (dashed).}
\label{fig3}
\end{figure}

For narrow rings
the quantum oscillations of the spin transmittance can be
compared with the analytical  result \cite{Malshukov1}
\begin{equation}\label{eq6}
\langle g_{yy} \rangle = \frac{-g_0 \kappa^2 \cos[\pi\sqrt{1+4(d/L_{so})^2}]
}{\kappa^2+4\sin^2\{\cos[\pi\sqrt{1+4(d/L_{so})^2}]\}},
\end{equation}
with the spin independent conductance $g_0$,
the averaged radius $d=\sqrt{(r^{2}_1 +r^{2}_2)/2}$,
and the parameter $\kappa=\sqrt{2T_w/\tau}$,
where $\tau$ denotes the mean escape time
and $T_w$ represents the average duration for one winding.
The equation (\ref{eq6}) is valid for $\kappa$ and $|r_2-r_1|/L_{so} \ll 1$.
The dashed curve in Fig. \ref{fig3} is rescaled from Fig. \ref{fig2} (c)
and is bounded by $\langle g_{yy}\rangle=0.56$,
which is the spin independent conductance given by Eq. (\ref{eq5})
with $\mbox{trace}\{\sigma_i S_\gamma \sigma_j S_\gamma^\dagger\}$
substituted for 2.
The solid curve in Fig. \ref{fig3} is obtained from Eq. (\ref{eq6})
with $g_0=0.56$.
The positions and magnitudes of the maxima and minima on both curves
almost coincide for first several oscillations,
but the deviation, as expected, increases for smaller $L_{so}$.
Furthermore, the dashed curve is much smoother than the solid one,
because the leads in the ring in Fig. \ref{fig2} (d) are relatively wide
and the mean escape time of the particle is not long enough to produce the
sharp interference pattern from many windings.
Simulations (not shown in Fig. \ref{fig3}) with narrower leads result
in curves with the sharp peaks resembling the analytical solid curve.
%%%%%%%%%%%%%%%%%%%%%%%%%%%
%%%%%%%%%%%%%%%%%%%%%%%%%%%

\begin{figure}[htbp!]
\includegraphics[width=8.5cm]{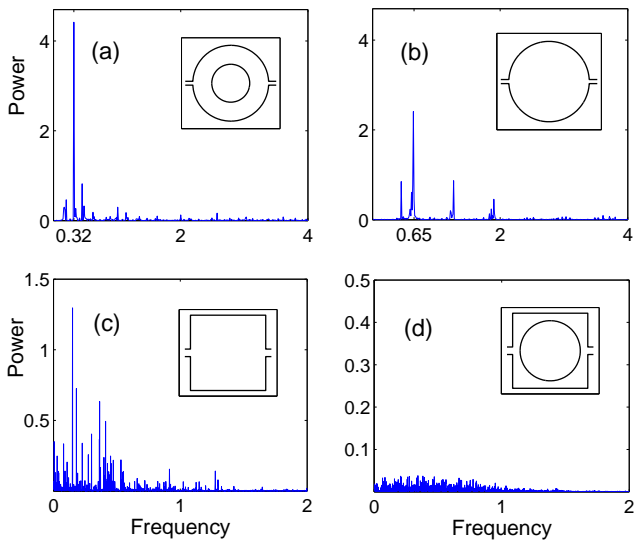}
\caption{
The power spectrum in (a) the annular billiard with $r_1=0.5$
and (b) the circle billiard with $r_1=0$, for
$(r_2,w,n_w,n_\theta,n_{L_{so}},L_{so1},L_{so2})$
=$(1,0.1,4,40,3000,0.01,10)$;
and in (c) the square billiard with $r_1=0$ and 
(d) Sinai billiard with $r_1=0.8$, for
$(r_2,w,n_w,n_\theta,n_{L_{so}},L_{so1},L_{so2})$
=$(1,0.2,4,100,20000,0.001,100)$.}
\label{fig4}
\end{figure}

\begin{figure}[htbp!]
\includegraphics[width=8.5cm]{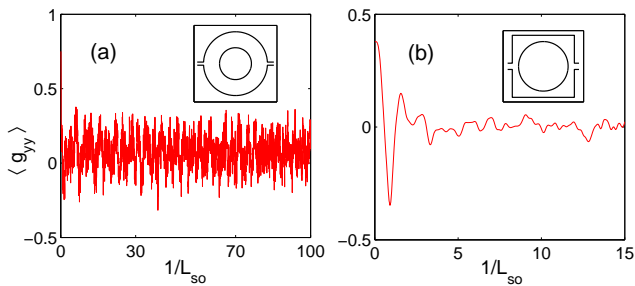}
\caption{
The conductance oscillation $\langle g_{yy}\rangle$ of
(a) the annular billiard from Fig. 4 (a) and
(b) the Sinai billiard from Fig. 4 (d).}
\label{fig5}
\end{figure}

An apparent trend seen in Fig. \ref{fig2} from (d) to (b) is a faster
damping of the dominating periodic oscillations with increasing ring width.
However, our calculations of the spin conductance in annular and
circle billiards gave an unexpected result,
in which new periodic oscillations emerge, as shown in
the power spectra in Fig. \ref{fig4} (a) and (b).
They are not so prominent as those in the thin rings,
nevertheless, are rather distinguishable.
The power spectrum for the annular billiard in Fig. \ref{fig4} (a)
consists of a dominating frequency around $0.32$ and several secondary peaks. 
The power spectrum for the circle billiard in Figure \ref{fig4} (b) shows that
the main  peak is still presents near $0.65$.
A simple analysis indicates that the trajectories contributing
to the main peak in Fig. \ref{fig4} (a) and (b) are those with
injection angles close to $0$.
Notably, the oscillations in the thin rings and those in the annular
and the circle billiards are fundamentally different.
In the former case, the oscillations with one dominating
frequency are of universal nature, because they can be observed
in arbitrary geometry, both in chaotic and regular systems,
as far as the billiards have the shape of a thin  loop \cite{Malshukov1}.
However, oscillations of the spin conductance in the circle and annular
billiards, like in Fig. \ref{fig5} (a), are entirely determined by the 
specific geometry of these systems.
%%%%%%%%%%%%%%%%%%%%%%%%%%%
%%%%%%%%%%%%%%%%%%%%%%%%%%%

Figure \ref{fig4} (c) shows the power spectrum for the square billiard.
Unlike the spectra of the ring and annular billiards, this spectrum
consists of a broad range of peaks with comparable intensities.
Figure \ref{fig4} (d) shows the power spectrum of a Sinai billiard,
which is a strongly chaotic system.
This spectrum is more uniformly distributed and less intense in
comparison to all other regular billiards studied above.
It is interesting to note that two apparent oscillations of the spin
conductance can be still observed in this chaotic billiard
[Fig. \ref{fig5} (b)], although its shape does not look like a thin loop.
%%%%%%%%%%%%%%%%%%%%%%%%%%%
%%%%%%%%%%%%%%%%%%%%%%%%%%%

As seen from above examples, the frequency spectra of different
billiards reflect their geometry.
For the annular and circle billiards in Fig. \ref{fig4} (a) and (b),
all segments $l_i$'s along a trajectory
are of equal length [Fig. \ref{fig1} (b)],
which gives the unique characteristic frequency of all individual
$S_{\gamma_i}$'s in Eq. (\ref{eq4}). When these operators
are multiplied in  Eq. (\ref{eq3}) and (\ref{eq5}), a descrete set of the 
power spectrum harmonics is generated. 
For the square billiard in Fig. \ref{fig4} (c),
the segments $l_i$'s along a trajectory have different lengths.
As a result, the broad range of frequencies contribute to the power spectrum. 
In the case of the Sinai billiard, the frequency distribution is 
more uniform compared to the spectra of the regular systems in Fig.
\ref{fig4} (a), (b), and (c).
Intuitively, this difference can be understood in the following way.
Each trajectory contributes to a finite set of frequencies
in the power spectrum.
In regular systems, the trajectories with close initial conditions 
stay close to each other during their liftime within a billiard
and form a trajectory bundle, which contributes to 
strong peaks in the power spectra.
However, these bundles are absent in chaotic systems,
because close trajectories exponentially fast move away from each other .

This work is supported by the National Science Council of the
R.O.C., Taiwan, under Contract No. NSC 90-2112-M-007-067,
the Royal Swedish and the Russian Academies of Sciences.
A.G. Mal'shukov acknowledges the hospitality of the National Center for
Theoretical Sciences, Taiwan.

\end{document}